\documentclass[%
reprint, superscriptaddress,
amsmath,amssymb,
aps,
]{revtex4-2}

\usepackage{graphicx}% Include figure files
\usepackage{dcolumn}% Align table columns on decimal point
\usepackage{bm}% bold math
\usepackage{xcolor}
\usepackage{physics}
\usepackage{cancel}

\begin{document}
	
	\preprint{APS/123-QED}
	
	\title{Berry bands and pseudo-spin of topological photonic phases}%
	
	\author{Samuel J.~Palmer}
	\email{samuel.palmer12@imperial.ac.uk}
	\affiliation{
		The Blackett Laboratory, Imperial College London, London, SW7 2AZ, UK
	}%
	
	\author{Vincenzo Giannini}%
	\homepage{http://www.GianniniLab.com}
	\affiliation{
		Instituto de Estructura de la Materia (IEM), Consejo Superior de Investigaciones Científicas (CSIC), Serrano 121, 28006, Madrid, Spain
	}%
	
	
	\begin{abstract}
		{
			Realising photonic analogues of the robust, unidirectional edge states of electronic topological insulators would improve our control of light on the nanoscale and revolutionise the performance of photonic devices. Here we show that new symmetry protected topological phases can be detected by reformulating energy eigenproblems as Berry curvature eigenproblems. The ``Berry bands'' span the same eigenspace as the original valence energy bands, but separate into pseudo-spinful and pseudo-spinless subspaces in $\mathrm{C}_2\mathcal{T}$-symmetric crystals. We demonstrate the method on the well-known case of Wu \& Hu~[Phys. Rev. Lett. 114, 223901 (2015)] and a recently discovered fragilely topological crystal, and show that both crystals belong to the same photonic analogue of the quantum spin-Hall effect. This work helps unite theory and numerics, and is useful in defining and identifying new symmetry-protected phases in photonics and electronics.
		}
	\end{abstract}
	
	\maketitle
	
	%\tableofcontents

	\emph{Introduction.---}
	When guiding light on the nanoscale, impurities, imperfections, and sharp corners can scatter light in unintended ways and limit the performance of photonic devices.
	This unintended scattering could be reduced if light can be guided using the robust, unidirectional states that arise at the surfaces of crystals with non-trivial band topologies. This is the one of the principal goals of topological nanophotonics \cite{rider2019perspective}. 
	Although non-trivial topological phases were first observed in the electronic bands of atomic crystals \cite{kane2005quantum,kane2005z,bernevig2006quantum,hasan2010colloquium,xiao2010berry,bansil2016colloquium}, photonic analogues of topological phases such as the quantum Hall effect (QHE) \cite{klitzing1980new,thouless1982quantized} and symmetry-protected phases such as the quantum spin-Hall effect (QSHE) \cite{kane2005z,kane2005quantum,bernevig2006quantum} have been built using photonic crystals: periodic nanostructures with tunable photonic bands \cite{john1987strong,yablonovitch1987inhibited}.
	
	The photonic QHE has robust surface states, but requires time-reversal symmetry to be broken \cite{haldane2008possible,raghu2008analogs,wang2009observation,poo2011experimental}. This can also be achieved in photonics using, for example, an external magnetic field~\cite{Mittal2019}, but in practice the time-reversal breaking responses of common materials are weak in the visible spectrum \cite{landau1982theoretical,rider2019perspective}. As such, there is a particular interest in photonic analogues of topological phases that are time-reversal symmetric, such as the QSHE.
	
	The QSHE can be considered as two counterpropagating instances of the QHE, one for each spin and with opposite magnetic fields to maintain time-reversal symmetry. In general, the surface states are not robust as crossings between the counterpropagating surface states can be gapped by non-spin-preserving perturbations \cite{fu2006time,ozawa2019topological}. However, with fermionic time-reversal symmetry, $\mathcal{T}^2=-1$, there are protected Kramers' degeneracies at time-reversal invariant momenta and the number of topological surface states propagating in each direction can only change by an even number \cite{asboth2016short}. The QSHE is therefore a $\mathbb{Z}_2$ topological phase, with either an even (trivial) or odd (non-trivial) number of edge states propagating in each direction \cite{kane2005z}.
	
\begin{figure}
		\includegraphics[width=\linewidth]{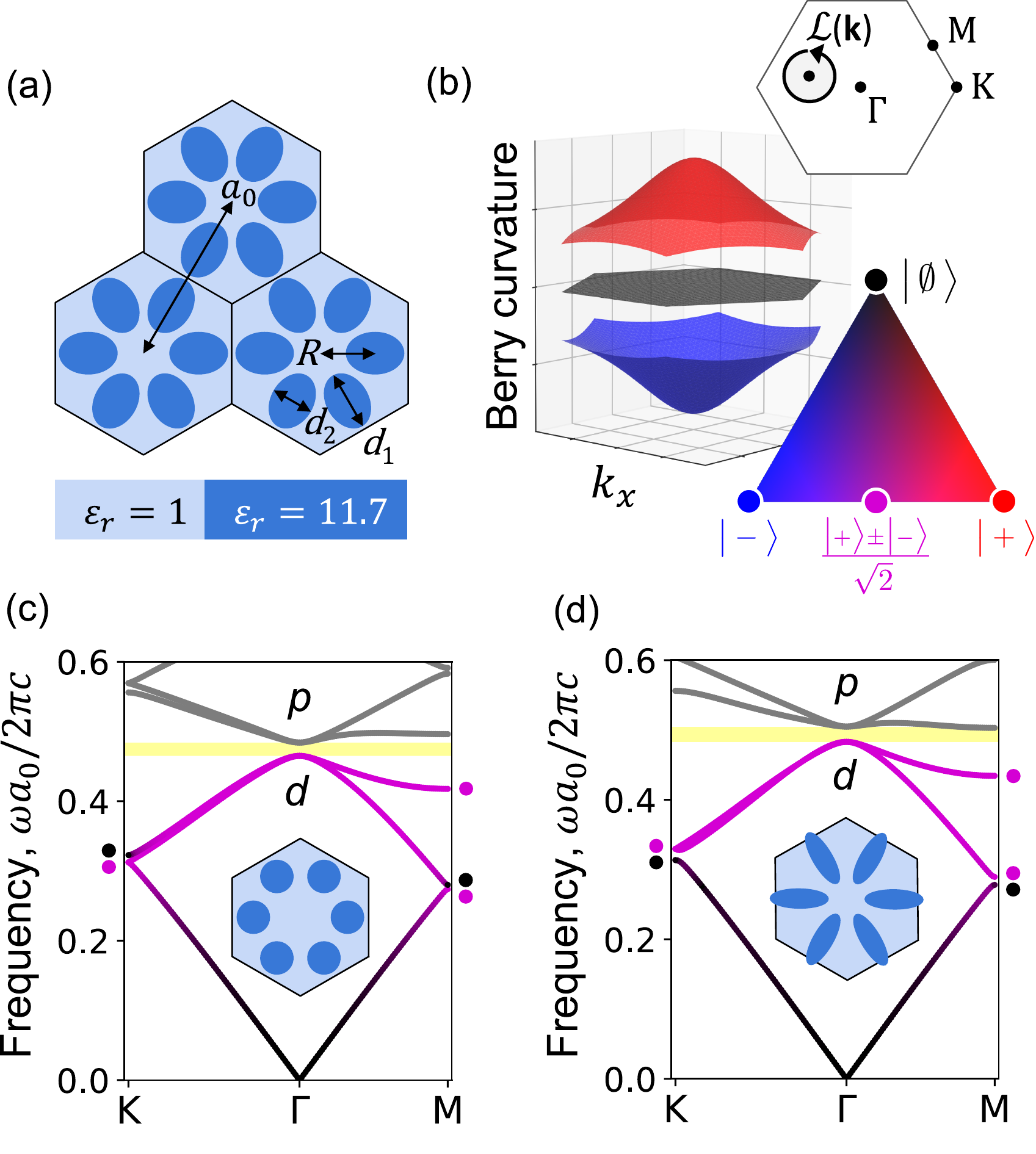}
		\caption{\label{fig:intro}(a) A topological photonic crystal consisting of dielectric cylinders ($\epsilon_r=11.7$) in air ($\epsilon_r=1$). The ellipsoidal cylinders have axes of lengths $d_1$ and $d_2$ and are arranged in rings of radius $R$ on a triangular lattice of site-to-site separation $a_0$. (b) In $\mathrm{C}_2\mathcal{T}$ symmetric crystals, the Berry curvature of the valence bands comes in zero (pseudo-spinless) or positive/negative (pseudo-spinful) pairs. (c) TM polarised bands of a topological photonic crystal with $a_0/R=3.125$, $d_1=d_2=2R/3$ and (d) a similar crystal with $a_0/R=3$, $d_1=0.4$, $d_2=0.13$. We show that there is a non-trivial $\mathbb{Z}_2$ topological phase for both crystals that can be observed in the Wilson loops of the pseudo-spinful subspace. Previous works had shown that Wilson loops through the full valence band spaces fail to detect a non-trivial topological phase \cite{blanco2019engineering,blanco2020tutorial}. By colouring the bands according to their pseudo-spin composition, we see that this is because the valence band spaces contain a mixture of pseudo-spinful and pseudo-spinless states.}
	\end{figure}

	An elegant photonic analogue of the QSHE was proposed by Wu and Hu \cite{wu2015scheme} where the circular polarisation of light mimics the spin space of the electrons, and crystalline symmetries produce a fermionic pseudo-time-reversal symmetry that protects the edge states at $\Gamma$. The design consists of hexagonal rings of cylinders arranged on a triangular lattice, as shown in Fig.~\ref{fig:intro}a. When the cylinders are circular ($d_1=d_2$) there is a certain ring radius ($a_0/R=3$) where the cylinders form a honeycomb lattice and the transverse magnetic modes meet at a double Dirac point between $p$ (dipolar) and $d$ (quadrupolar) modes at $\Gamma$. Breaking the Dirac point by expanding the rings of cylinders produces an effective Hamiltonian that is equivalent (in the vicinity of $\Gamma$) to the Bernevig-Hughes-Zhang model of the QSHE \cite{bernevig2006quantum}.
	Thanks to its simplicity, this model has been widely studied recently \cite{blanco2019engineering,Barik2016a,smirnova2019third,Parappurath2020,Liu2020,orazbayev2019quantitative,proctor2019exciting} and has been used to show the importance of finite size effects in topological photonics and the emergence of topological particle resonances or topological whispering gallery modes~\cite{Siroki2017,Yang2018}.
	
	It has been shown that in some cases \cite{Pocock2018,Pocock2019}, identifying topological photonic phases requires proper consideration of long range interactions and retardation. However, to our knowledge the expected topological indices have not been determined from full-wave calculations over the full Brillouin zone for the structures proposed by Wu and Hu, but only with approximations near the $\Gamma$ point \cite{wu2015scheme,Barik2016a,smirnova2019third,Liu2020}. This has led to the misinterpretation of the protecting symmetry and to the belief that this crystal is not a photonic analogue of the QSHE~\cite{blanco2019engineering,orazbayev2019quantitative}.

	In this letter we show how the bands of Berry curvature, which we call the ``Berry bands'', form a natural basis for the pseudo-spin of a crystal, as shown in Fig.~\ref{fig:intro}b, and how symmetry-protected phases can be identified using Wilson loops \cite{asboth2016short,vanderbilt2018berry} through the pseudo-spin subspaces. The calculations are performed with \texttt{Peacock.jl}, a freely available Julia package for studying topological photonics using the plane-wave expansion method and Wilson loops~\cite{Pea}. Taking, for example, the well known crystal of Wu and Hu shown in Fig.~\ref{fig:intro}c, we show that the spectra of Wilson loops applied directly to the energy bands imply the existence of corner states but not necessarily chiral edge states \cite{po2018fragile,ahn2019stiefel}, whereas taking Wilson loops of the Berry bands reveals the photonic analogy to the QSHE. We also show that a related photonic crystal whose energy bands have fragile topology \cite{blanco2019engineering,blanco2020tutorial}, shown in Fig.~\ref{fig:intro}d, belongs to the same symmetry-protected phase as the crystal of Wu and Hu \cite{wu2015scheme}. By colouring the energy bands of Figs.~\ref{fig:intro}c\nobreakdash-d according to their pseudo-spin composition, we see that the fragile topology arises when the energy bands are gapped in such a way as to separate the pseudo-spinful and pseudo-spinless spaces. These results help unite theory and numerics, and may be useful in defining and identifying new symmetry-protected phases.

	\emph{\\Gapped topological phases.---}Two gapped Hamiltonians are in different topological phases when it is impossible to adiabatically deform from one to the other without closing the energy gap. In 2D, the topological index of each phase is the total Chern number \cite{raghu2008analogs} of the valence band space, $\mathcal{H}_\mathrm{val}$. Later, we will discuss how $C_\mathrm{val}$ can be calculated using Wilson loops. In the QHE, unidirectional surface states are observed at the interface between two different topological phases. The net number of edge states travelling in a certain direction is the difference in Chern number of the two phases, $\Delta C_\mathrm{val}$.
	
	Moreover, two Hamiltonians belong to different \emph{symmetry-protected} phases when we may adiabatically deform from one to the other without closing the energy gap, but only if the protecting symmetry is necessarily broken during the deformation. Systems in different symmetry-protected phases must belong to the same general topological phase and cannot be distinguished by $C_\mathrm{val}$. For example, both the trivial and non-trivial phases in the QSHE have $C_\mathrm{val}=0$. Instead, the valence band space must be decomposed into subspaces, $\mathcal{H}_\mathrm{val}(\boldsymbol{k}) = \bigoplus_{n=1}^N \mathcal{H}_n(\boldsymbol{k})$. If the projectors onto each subspace are smooth and periodic throughout the Brillouin zone, then each subspace has a well-defined Chern number, \cite{gresch2017z2pack}, $C_\mathrm{val} = \sum_{n=1}^N C_n$. 
	
	While there are many ways of decomposing $\mathcal{H}_\mathrm{val}$ into subspaces, leading to different $\{C_n\}$, the topological indices of a symmetry-protected phase must be robust against perturbations unless the protecting symmetry is broken or the valence-conduction band gap is closed. One approach is to decompose the band space according to symmetries that (block) diagonalise the Hamiltonian \cite{gresch2017z2pack}. In the following sections we introduce ``bands of Berry curvature'', which we call the Berry bands, and explain how decomposing $\mathcal{H}_\mathrm{val}$ according to these Berry bands reveals topological phases in $\mathrm{C}_2\mathcal{T}$ symmetric photonic crystals that emulate spin using circularly polarised light.

	\emph{\\Wilson loops.---}
	We use operators known as Wilson loops for two purposes: first to decompose $\mathcal{H}_\mathrm{val}$ into subspaces according to the local Berry curvature, and then to calculate the corresponding topological indices of each subspace. The Wilson loop of a closed path, $\mathcal{L}$, is
	\begin{equation}
	\hat{W}_\mathcal{L}^{\{n\}} = \hat{P}(\boldsymbol{k}_1)\hat{P}(\boldsymbol{k}_N)\dots\hat{P}(\boldsymbol{k}_2)\hat{P}(\boldsymbol{k}_1),
	\end{equation}
	where $\hat{P}(\boldsymbol{k}_i) = \sum_{n \in \{n\}} \ket{u_n(\boldsymbol{k}_i)}\bra{u_n(\boldsymbol{k}_i)}$ are projectors onto the subspace of interest, and $\boldsymbol{k}_i$ are closely spaced points along $\mathcal{L}$. The action of the Wilson loop is to parallel transport a mode through this subspace. In general this produces a unitary mixing,
	\begin{equation}
	\hat{W}_\mathcal{L}^{\{n\}} \ket{u_i(\boldsymbol{k}_1)} = \sum_{j}U_{ij} \ket{u_j(\boldsymbol{k}_1)},
	\end{equation}
	but there exist eigenmodes for each Wilson loop that will each accumulate a gauge-invariant geometric phase known as a Berry phase, $\gamma_i$, without mixing \cite{vanderbilt2018berry},
	\begin{equation}
	\hat{W}_\mathcal{L}^{\{n\}} \ket{\tilde{u}_i(\boldsymbol{k}_1)} = \exp(i\gamma_i) \ket{\tilde{u}_i(\boldsymbol{k}_1)},
	\end{equation}
	where $\ket{\tilde{u}_i(\boldsymbol{k}_1)} = \sum_{j}V_{ij} \ket{u_j(\boldsymbol{k}_1)}$ and $\mathbf{V}$ is a unitary matrix that diagonalises $\mathbf{U}$ as $(\mathbf{V}^\dagger\mathbf{U}\mathbf{V})_{ij} = \delta_{ij}\exp(i\gamma_i)$.

	\emph{\\Berry bands and pseudo-spin.---}
	We generate a pseudo-spin basis for $\mathrm{C}_2\mathcal{T}$-symmetric crystals using Wilson loops around infinitesimally small paths $\mathcal{L}(\boldsymbol{k})$ enclosing $\boldsymbol{k}$, as shown in the inset of Fig.~\ref{fig:intro}b. We want the decomposition to be unaffected by perturbations that open or close energy gaps within the valence bands, so we build these Wilson loops using projectors onto the full valence band space,
	\[
	\hat{W}_{\mathcal{L}}^{\mathrm{val}}=\hat{P}_{\mathrm{val}}(\boldsymbol{k}_{1})\hat{P}_{\mathrm{val}}(\boldsymbol{k}_{N})\dots\hat{P}_{\mathrm{val}}(\boldsymbol{k}_{2})\hat{P}_{\mathrm{val}}(\boldsymbol{k}_{1}),
	\]
	thereby ensuring that different realisations of the same symmetry-protected phase, such as those in Figs.~\ref{fig:intro}c-d, are treated equally.
	
	\begin{figure}
		\centering
		\includegraphics[width=\linewidth]{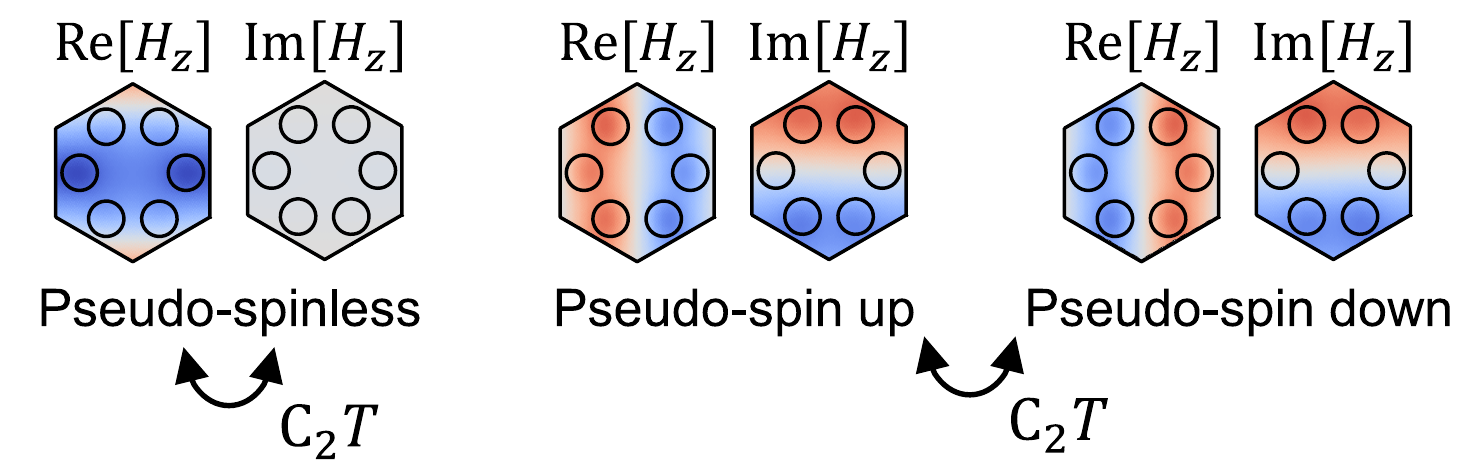}
		\caption{Pseudo-spin modes at $M$ for the topological crystal of Wu and Hu ($a_0/R=3.125, d_1=d_2=2R/3$). In $\mathrm{C}_2\mathcal{T}$-symmetric crystals the Berry curvature, $\mathcal{F}(\boldsymbol{k})$, comes in positive/negative pairs or is zero. The pseudo-spinless mode, $\mathcal{F}(\boldsymbol{k})=0$, is mapped to itself by $\mathrm{C}_2\mathcal{T}$ symmetry, whereas the pseudo-spin up and down modes, $\pm\mathcal{F}(\boldsymbol{k})$, are mapped to each other by $\mathrm{C}_2\mathcal{T}$ symmetry.}
		\label{fig:pseudo-spin}
	\end{figure}

	Because the Wilson loops are unitary operators, we can form a Hermitian eigenvalue problem for the non-Abelian Berry curvature, $\mathcal{F}_i(\boldsymbol{k})$,
	\begin{equation}
	\hat{H}_{\mathcal{F}}(\boldsymbol{k})\ket{\tilde{u}_{i}(\boldsymbol{k})}=\mathcal{F}_{i}(\boldsymbol{k})\ket{\tilde{u}_{i}(\boldsymbol{k})},\label{eq:HF}
	\end{equation}
	where $\hat{H}_{\mathcal{F}}(\boldsymbol{k})=\lim_{A\rightarrow0}[-i\log\hat{W}_{\mathcal{L}}^{\mathrm{val}}/A]$ and $\mathcal{F}_{i}(\boldsymbol{k})=\lim_{A\to0}[\gamma_{i}(\boldsymbol{k})/A]$ for a vanishingly small loop $\mathcal{L}$ of area $A$ enclosing $\boldsymbol{k}$. This transforms the original energy eigenvalue problem, $E(\boldsymbol{k})$ to a ``Berry band'' eigenvalue problem, $\mathcal{F}(\boldsymbol{k})$, where the Berry bands span the valence band space and, as the Wilson loops are vanishingly small, inherit the spatial symmetries of the original energy bands, as shown in Fig.~\ref{fig:pseudo-spin}.
	
	The combined $\mathrm{C}_{2}$ and $\mathcal{T}$ symmetries act on the Berry bands as \cite{bradlyn2019disconnected,bouhon2019wilson}
	\begin{align}
		\mathrm{C}_{2}\mathcal{T}\,\hat{H}_{\mathcal{F}}(\boldsymbol{k})\,(\mathrm{C}_{2}\mathcal{T})^{-1} & =-\hat{H}_{\mathcal{F}}(\boldsymbol{k}),\label{eq:C2T_HF}
	\end{align}
	and therefore each Berry band in a $\mathrm{C}_2\mathcal{T}$-symmetric crystal either has a value of zero (pseudo-spinless) or is part of a positive/negative pair (pseudo-spinful). The pseudo-spinful Berry bands are circularly polarised,
	\begin{align}
		\ket{\tilde{u}_{\pm}(\boldsymbol{k})} & =\frac{1}{\sqrt{2}}\ket{u_{1}(\boldsymbol{k})}\pm\frac{i}{\sqrt{2}}\ket{u_{2}(\boldsymbol{k})},
	\end{align}
	where $\mathrm{C}_2\mathcal{T}\ket{\tilde{u}_{\pm}(\boldsymbol{k})}=\ket{\tilde{u}_{\mp}(\boldsymbol{k})}$, and $\ket{u_1(\boldsymbol{k})}$ and $\ket{u_2(\boldsymbol{k})}$ are invariant under $\mathrm{C}_2\mathcal{T}$.

\begin{figure}
		\includegraphics[width=\linewidth]{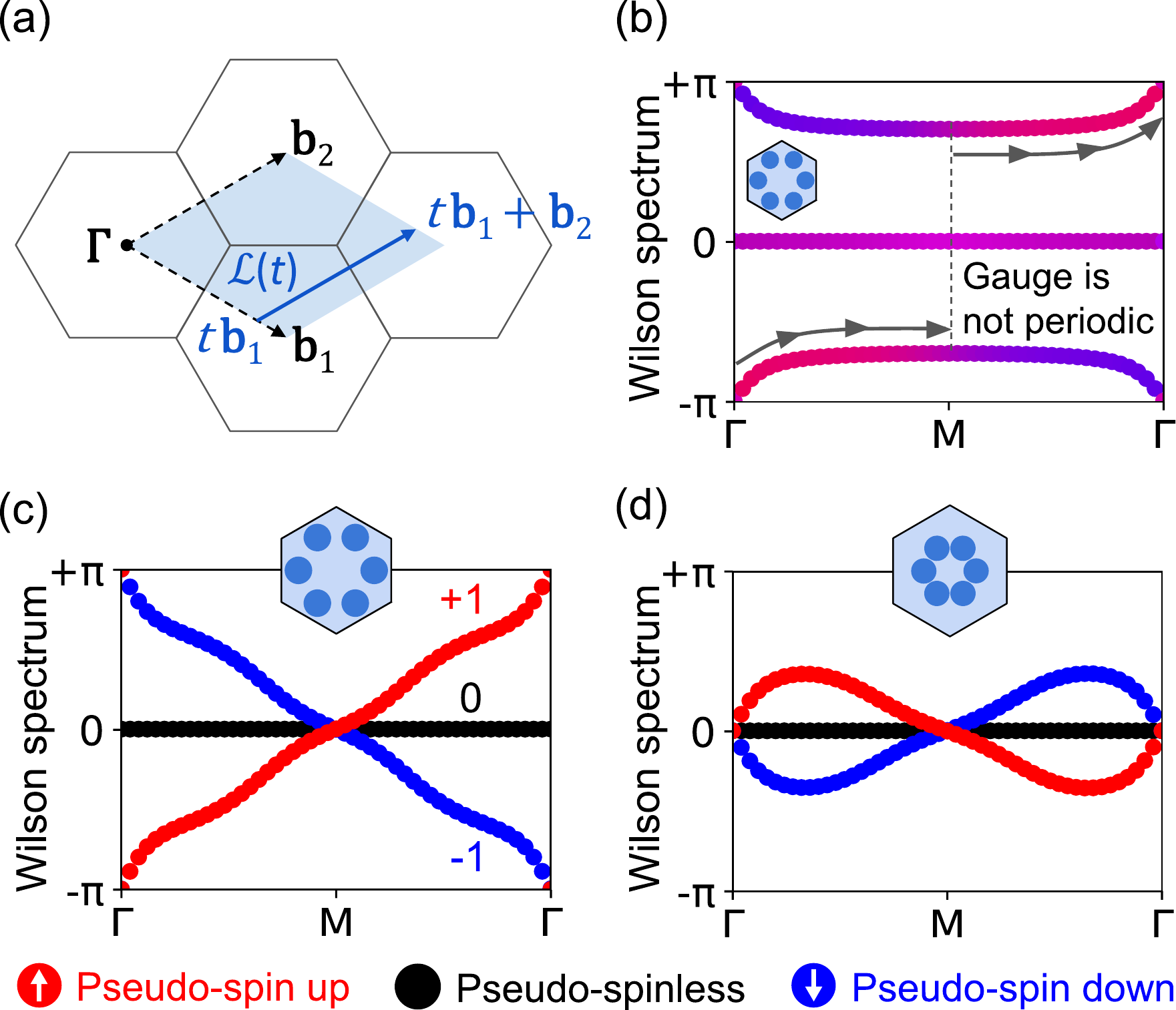}
		\caption{\label{fig:application}
			(a) The Chern number of a space can be observed as a winding in the spectrum of Wilson loops made on a series of paths, $\mathcal{L}(t)$. These paths sweep the Brillouin zone (shaded blue region) as $t \to t+1$.
			(b) For the topological crystal of Wu and Hu ($a_0/R=3$, $d_1=d_2=2R/3$) the Wilson loop spectrum of the total valence band space has zero total winding, as expected from time-reversal symmetry. By colouring the spectrum according to the pseudo-spin composition, we see that the individual windings are not well defined as the resultant gauge is not smooth and periodic. No spin-Chern numbers are observed.
			(c) In contrast, Wilson loops made through the pseudo-spinful subspace (Berry bands) present well-defined non-trivial windings. The corresponding $\mathbb{Z}_2$ index is shown to be protected by $\mathrm{C}_2\mathcal{T}$ symmetry.
			(d) Similarly, the pseudo-spin winding of a crystal without band inversion, $a_0/R=2.9$, $d_1=d_2=2R/3$, has a trivial winding protected by $\mathrm{C}_2\mathcal{T}$ symmetry.}
	\end{figure}

	\emph{\\$\mathrm{C}_2\mathcal{T}$-protected $\mathbb{Z}_2$ phase.---}
	The total Chern number of the valence bands is equivalent \cite{blanco2020tutorial} to the total spectral winding of Wilson loops built from projectors onto the valence bands. These Wilson loops are made along a series of parallel paths $\mathcal{L}(t)$, shown in Fig.~\ref{fig:application}a, where the paths sweep the Brillouin zone as $t \to t+1$. Fig.~\ref{fig:application}b shows the Wilson loop spectra of the three-dimensional valence band space for the topological crystal introduced by Wu and Hu \cite{wu2015scheme}. The total winding and therefore $C_\mathrm{val}$ are both zero, as expected for the valence bands of a time-reversal symmetric system \cite{vanderbilt2018berry}. However, colouring the Wilson loop spectra according to their pseudo-spin composition reveals that the windings of the individual eigenvalues are not smooth and periodic, and the individual spin-Chern numbers cannot be determined from this analysis.

	One way of proceeding is to consider the two eigenvalues of Fig.~\ref{fig:application}b that cross $\pm\pi$ an odd number of times, indicating a non-trivial 2D Stiefel-Whitney insulating phase expected to host corner states \cite{ahn2019stiefel}. This is also known as an obstructed atomic limit (OAL) \cite{blanco2019engineering} because the pair Wilson loop eigenvalues averaging around $\pm\pi$ indicate that two of the maximally localised Wannier functions of this crystal are localised at the edges of the unit cell \cite{gresch2017z2pack,blanco2019engineering}. The Stiefel-Whitney insulator/OAL is expected to host topological corner states \cite{ahn2019stiefel,po2018fragile}, but the observation of chiral edge states \cite{wu2015scheme,orazbayev2019quantitative,Liu2020,Parappurath2020} remain unexplained by this analysis.

	Alternatively, we may use the Berry bands introduced in the previous section to decompose the valence band space into a pseudo-spinless subspace and a pseudo-spinful subspace,
	\begin{equation}
	\mathcal{H}_\mathrm{val} = \mathcal{H}_\emptyset \oplus \mathcal{H}_\pm.
	\end{equation}
	Fig.~\ref{fig:application}c shows the spectra of Wilson loops made separately through $\mathcal{H}_\emptyset$ and $\mathcal{H}_\pm$ for the same crystal as in Fig.~\ref{fig:application}b. There is no mixing of the pseudo-spinful states for Wilson loops through a 2-band pseudo-spinful space, as shown in the supplementary material. The individual spectra are smooth and periodic, and so the corresponding spin-Chern numbers, $\{C_-,C_\emptyset,C_+\}=\{-1,0,+1\}$, are well defined. In other words, the valence bands consist of two QHE related to each other by $\mathrm{C}_2\mathcal{T}$ symmetry and an additional trivial subspace that is separable via the Berry bands. With pseudo-Kramers degeneracy enforced at $\Gamma$ by $\mathrm{C}_6$ symmetry \cite{wu2015scheme}, we conclude that the crystal is a photonic analogue of the QSHE, with the parity of $C_+$ and $C_-$ as the $\mathbb{Z}_2$ topological index. For the corresponding trivial phase, the spin-Chern numbers are $\lbrace C_n \rbrace = \lbrace 0,0,0 \rbrace$, as indicated by the lack of winding in Fig.~\ref{fig:application}d.

	\begin{figure}
		\includegraphics[width=\linewidth]{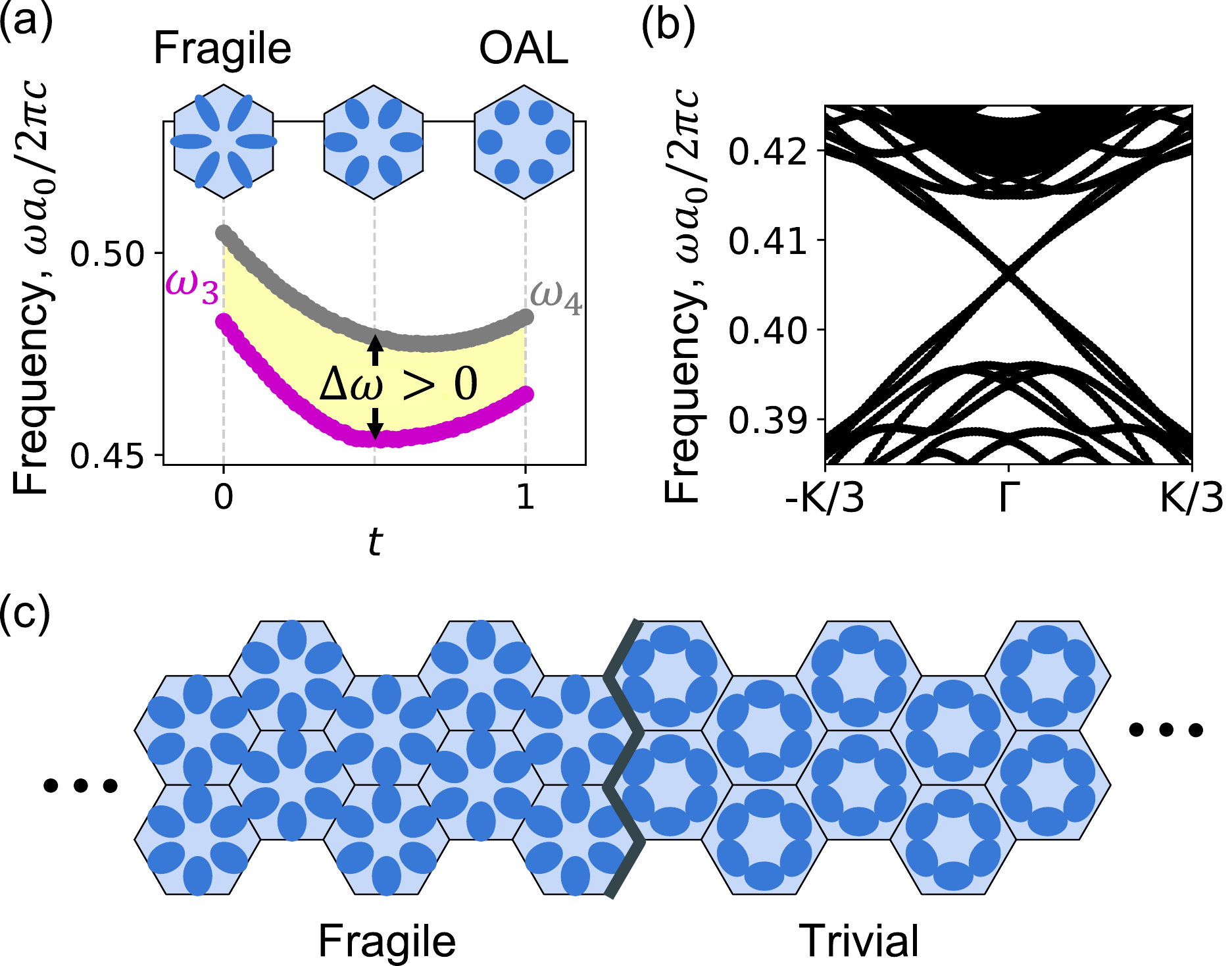}
		\caption{\label{fig:supporting-evidence}
			(a) Band gap for a continuously deformed crystal. It is possible to adiabatically deform between the obstructed atomic limit ($a_0/R{=}2.9$, $d_1{=}d_2{=}2R/3$) and fragilely topological ($a_0/R{=}3, d_1{=}0.4a_0, d_2{=}0.13a_0$) crystals without breaking $\mathrm{C}_2\mathcal{T}$ symmetry or closing the topological band gap, $\Delta\omega =\omega_4-\omega_3 > 0$. This means both crystals must belong to the same $\mathrm{C}_2\mathcal{T}$-protected phase.
			(b)-(c) Edge states are observed at the interface between a fragilely topological ($a_0/R{=}3, d_1{=}0.35a_0, d_2{=}0.25a_0$) and a trivial ($a_0/R{=}3, d_1{=}0.25a_0, d_2{=}0.35a_0$) crystal. The parameters were chosen so the two crystals had an overlapping band gap at $\omega a_0/2 \pi c \approx 0.4$. The emergence of these edge states agrees with the measured $\mathrm{C}_2\mathcal{T}$-protected $\mathbb{Z}_2$ indices of the pseudo-spin subspaces.}
	\end{figure}
	
	We also studied the ``fragilely topological'' crystal introduced by Blanco de Paz et al \cite{blanco2019engineering}, and found the same spectral winding as seen in Fig.~\ref{fig:application}c, indicating that both crystals are photonic analogues of the QSHE. This is supported by Fig.~\ref{fig:supporting-evidence}a which shows that it is possible to adiabatically deform between the crystals without closing the topological band gap or breaking $\mathrm{C}_2\mathcal{T}$ symmetry, meaning the two crystals are in the same $\mathrm{C}_2\mathcal{T}$-protected phase, and also by Figs.~\ref{fig:supporting-evidence}b\nobreakdash-c which show topological edge states at the interface between the trivial and fragilely topological crystals. Here, as in other works \cite{wu2015scheme,orazbayev2019quantitative,proctor2019exciting}, there is a small gap in the edge modes (around 3\% of the bulk valence-conduction gap) as the presence of the interface is a $\mathrm{C}_6$ breaking perturbation that lifts the pseudo-Kramers degeneracy. See the supplementary material for more detailed band diagrams of the adiabatic deformation of Fig.~\ref{fig:supporting-evidence}a.

	\emph{\\Conclusion.---}
	We show that new symmetry protected topological phases can be identified by reformulating the energy eigenvalue problem as a Berry curvature eigenproblem. In $\mathrm{C}_2\mathcal{T}$-symmetric crystals the ``Berry bands'' separate into pseudo-spinless (linearly polarised) and pseudo-spinful (circularly polarised) subspaces. Using straight Wilson loops through the pseudo-spinful subspaces of crystals with three valence bands detects topological phases where Wilson loops through the energy valence bands fail to do so. We demonstrate the method on the well-known photonic crystal of Wu and Hu and a recently discovered `fragilely topological' crystal and show that both crystals belong to the same $\mathrm{C}_2\mathcal{T}$-protected $\mathbb{Z}_2$ topological phase that emulates the quantum-spin Hall effect in photonics. Studying the topology of photonic crystals with larger valence band spaces may be possible using bent Wilson loops \cite{bouhon2019wilson,bradlyn2019disconnected} through the pseudo-spinful subspaces. The method presented here helps unite the numerics and theory of photonic topological insulators, and could also be applied to find new symmetry-protected phases in electronic systems \cite{wu2016topological}.

	\begin{acknowledgments}
		S.J.P. acknowledges his studentship from the Centre for Doctoral Training on Theory and Simulation of Materials at Imperial College London funded by EPSRC Grant No. EP/L015579/1.
		V.G. acknowledges the Spanish Ministerio de Economia y Competitividad for financial support through the grant NANOTOPO (FIS2017-91413-EXP) and also the Ministerio de Ciencia, Innovació n y Universidades through the grant MELODIA (PGC2018-095777-B-C21).
	\end{acknowledgments}
	\bibliography{references}% Produces the bibliography via BibTeX.

\clearpage

\onecolumngrid
\appendix

\onecolumngrid

\section{Wilson loops through the two-band pseudo-spinful subspace commute in $\mathrm{C}_2\mathcal{T}$-symmetric crystals with three valence bands}

In this section we will assume that we have a $\mathrm{C}_2\mathcal{T}$-symmetric crystal with three valence bands and shall show that any Wilson loop through the pseudo-spinful subspace will not mix the pseudo-spinful states together. An arbitrary Wilson loop can be written in unitary matrix form as \cite{vanderbilt2018berry}
\begin{align}
	\left[\mathbf{W}_{\mathcal{L}}(\boldsymbol{k})\right]_{ij} & =\bra{u_{i}(\boldsymbol{k}_{1})}\prod_{n=1}^{N-1}\boldsymbol{\mathcal{M}}(\boldsymbol{k}_{n+1},\boldsymbol{k}_{n})\ket{u_{j}(\boldsymbol{k}_{1})}\label{eq:wilson-matrix}
\end{align}
where $\boldsymbol{k}_{n}$ are points along the closed loop $\mathcal{L}$, and $\boldsymbol{\mathcal{M}}(\boldsymbol{k}_{n+1},\boldsymbol{k}_{n})$ is the best unitary approximation of the overlap matrix between the relevant states at $\boldsymbol{k}_{n+1}$ and $\boldsymbol{k}_{n}$,
\begin{equation}
\left[\mathbf{M}(\boldsymbol{k}_{n+1},\boldsymbol{k}_{n})\right]_{ij}=\braket{u_{i}(\boldsymbol{k}_{n+1})}{u_{j}(\boldsymbol{k}_{n})}.
\end{equation}
The unitary approximation $\boldsymbol{\mathcal{M}}(\boldsymbol{k}_{n+1},\boldsymbol{k}_{n})=\mathbf{U}\mathbf{V}^{\dagger}$ is calculated using the singular value decomposition $\mathbf{M}(\boldsymbol{k}_{n+1},\boldsymbol{k}_{n})=\mathbf{U}\boldsymbol{\Lambda}\mathbf{V}^{\dagger}$ where $\mathbf{U}$ and $\mathbf{V}$ are unitary matrices. First we will determine the form of $\mathbf{M}(\boldsymbol{k}_{n+1},\boldsymbol{k}_{n})$, and then we will show that its best unitary approximation, $\boldsymbol{\mathcal{M}}(\boldsymbol{k}_{n+1},\boldsymbol{k}_{n})$, is diagonal.

The pseudo-spinful Berry bands can be written as $\ket{\tilde{u}_{\pm}(\boldsymbol{k})}=\frac{1}{\sqrt{2}}\ket{u_{1}(\boldsymbol{k})}\pm\frac{i}{\sqrt{2}}\ket{u_{2}(\boldsymbol{k})}$ where $\ket{u_{1}(\boldsymbol{k})}$ and $\ket{u_{2}(\boldsymbol{k})}$ are a $\mathrm{C}_2\mathcal{T}$-invariant basis. The first element of the overlap matrix is
\begin{align}
	\braket{u_{+}(\boldsymbol{k}_{n+1})}{u_{+}(\boldsymbol{k}_{n})} & =\int\left(\frac{1}{\sqrt{2}}u_{1}^{*}(\boldsymbol{r},\boldsymbol{k}_{n+1})-\frac{i}{\sqrt{2}}u_{2}^{*}(\boldsymbol{r},\boldsymbol{k}_{1})\right)\nonumber \\
	& \quad\quad\quad\quad\cdot\left(\frac{1}{\sqrt{2}}u_{1}(\boldsymbol{r},\boldsymbol{k}_{2})+\frac{i}{\sqrt{2}}u_{2}(\boldsymbol{r},\boldsymbol{k}_{2})\right)\,\mathrm{d}^{2}r\\
	& =\frac{1}{2}\alpha_{11}+\frac{i}{2}\alpha_{12}-\frac{i}{2}\alpha_{21}+\frac{1}{2}\alpha_{22},
\end{align}
where $\alpha_{ij}(\boldsymbol{k}_{n+1},\boldsymbol{k}_{n})=\int u_{i}^{*}(\boldsymbol{r};\boldsymbol{k}_{n+1})u_{j}(\boldsymbol{r};\boldsymbol{k}_{n})\,\mathrm{d}^{2}r$. We can show that each $\alpha_{ij}$ is real by separating $u_{i}(\boldsymbol{r},\boldsymbol{k})=f_{i}(\boldsymbol{r},\boldsymbol{k})+ig_{i}(\boldsymbol{r},\boldsymbol{k})$ and recognising that the $\mathrm{C}_2\mathcal{T}$ symmetry of $u_{i}(\boldsymbol{r},\boldsymbol{k})$ requires that $f_{i}(\boldsymbol{r},\boldsymbol{k})$ and $g_{i}(\boldsymbol{r},\boldsymbol{k})$ are even and odd functions of $\boldsymbol{r}$, respectively. Therefore 
\begin{align}
	\alpha_{ij}(\boldsymbol{k}_{n+1},\boldsymbol{k}_{n}) & =\int\Bigl(f_{i}(\boldsymbol{r},\boldsymbol{k}_{n+1})-ig_{i}(\boldsymbol{r},\boldsymbol{k}_{n+1})\Bigr)\cdot\Bigl(f_{j}(\boldsymbol{r},\boldsymbol{k}_{n})+ig_{j}(\boldsymbol{r},\boldsymbol{k}_{n})\Bigr)\,\mathrm{d}^{2}r\\
	& =\int\Bigl(f_{i}(\boldsymbol{r},\boldsymbol{k}_{n+1})f_{j}(\boldsymbol{r},\boldsymbol{k}_{n})+g_{i}(\boldsymbol{r},\boldsymbol{k}_{n+1})g_{j}(\boldsymbol{r},\boldsymbol{k}_{n})\Bigr)\,\mathrm{d}^{2}r\\
	& \quad\quad+i\cancelto{0}{\int f_{i}(\boldsymbol{r},\boldsymbol{k}_{n+1})g_{j}(\boldsymbol{r},\boldsymbol{k}_{n})\,\mathrm{d}^{2}r}-i\cancelto{0}{\int g_{i}(\boldsymbol{r},\boldsymbol{k}_{n+1})f_{j}(\boldsymbol{r},\boldsymbol{k}_{n})\,\mathrm{d}^{2}r},\nonumber 
\end{align}
where the last two terms integrate to zero because the integrands are odd.

Therefore 
\begin{align}
	\braket{u_{+}(\boldsymbol{k}_{n+1})}{u_{+}(\boldsymbol{k}_{n})} & =\frac{\alpha_{11}+\alpha_{22}}{2}+i\frac{\alpha_{12}-\alpha_{21}}{2}
\end{align}
and similarly,
\begin{align}
	\braket{u_{-}(\boldsymbol{k}_{n+1})}{u_{-}(\boldsymbol{k}_{n})} & =\frac{\alpha_{11}+\alpha_{22}}{2}-i\frac{\alpha_{12}-\alpha_{21}}{2},\\
	\braket{u_{+}(\boldsymbol{k}_{n+1})}{u_{-}(\boldsymbol{k}_{n})} & =\frac{\alpha_{11}-\alpha_{22}}{2}-i\frac{\alpha_{12}+\alpha_{21}}{2},\\
	\braket{u_{-}(\boldsymbol{k}_{n+1})}{u_{+}(\boldsymbol{k}_{n})} & =\frac{\alpha_{11}-\alpha_{22}}{2}+i\frac{\alpha_{12}+\alpha_{21}}{2},
\end{align}
such that on the basis of $\ket{u_\pm(\boldsymbol{k}}$, the overlap matrix is
\begin{equation}
\mathbf{M}(\boldsymbol{k}_{n+1},\boldsymbol{k}_{n})=\left[\begin{array}{cc}
\braket{u_{+}(\boldsymbol{k}_{n+1})}{u_{+}(\boldsymbol{k}_{n})} & \braket{u_{+}(\boldsymbol{k}_{n+1})}{u_{-}(\boldsymbol{k}_{n})}\\
\braket{u_{-}(\boldsymbol{k}_{n+1})}{u_{+}(\boldsymbol{k}_{n})} & \braket{u_{-}(\boldsymbol{k}_{n+1})}{u_{-}(\boldsymbol{k}_{n})}
\end{array}\right]=\left[\begin{array}{cc}
a & b\\
b^{*} & a^{*}
\end{array}\right]
\end{equation}
where $a=\frac{1}{2}(\alpha_{11}+\alpha_{22})+\frac{i}{2}(\alpha_{12}-\alpha_{21})$ and $b=\frac{1}{2}(\alpha_{11}-\alpha_{22})-\frac{i}{2}(\alpha_{12}+\alpha_{21})$.

From the singular value decomposition, $\mathbf{M}(\boldsymbol{k}_{n+1},\boldsymbol{k}_{n})=\mathbf{U}\boldsymbol{\Lambda}\mathbf{V}^{\dagger},$ the unitary part of the matrix is

\begin{equation}
\boldsymbol{\mathcal{M}}(\boldsymbol{k}_{n+1},\boldsymbol{k}_{n})=\mathbf{U}\mathbf{V}^{\dagger}=\left[\begin{array}{cc}
\frac{e^{i\phi}}{2}\left[1+\Delta\right] & \frac{e^{i\theta}}{2}\left[1-\Delta\right]\\
\frac{e^{-i\theta}}{2}\left[1-\Delta\right] & \frac{e^{-i\phi}}{2}\left[1+\Delta\right]
\end{array}\right].
\end{equation}
where $a=\left|a\right|e^{\phi}$, $b=\left|b\right|e^{i\theta}$, $\Delta=\frac{\left|a\right|-\left|b\right|}{\left|\left|a\right|-\left|b\right|\right|}=\mathrm{sign}(\left|a\right|-\left|b\right|)$. However, we know that $\left|a\right|>\left|b\right|$ as our $\ket{u_{\pm}(\boldsymbol{k})}$ are smooth and continuous and continuous functions of $\boldsymbol{k}$. Therefore
\begin{equation}
\boldsymbol{\mathcal{M}}(\boldsymbol{k}_{n+1},\boldsymbol{k}_{n})=\left[\begin{array}{cc}
e^{i\phi} & 0\\
0 & e^{-i\phi}
\end{array}\right]\label{eq:diag-overlaps}
\end{equation}
is a diagonal matrix on the basis of $\ket{u_{\pm}}$.

Inserting Eq.~(\ref{eq:diag-overlaps}) into Eq.~(\ref{eq:wilson-matrix}), we see that the Wilson loops are diagonalised on the basis of $\ket{u_{\pm}(\boldsymbol{k})}$.

\section{\label{app:fragile-OAL-deformation}Deforming between fragilely topological and obstructed atomic limit crystals}

\begin{figure}[h!]
	\centering
	\includegraphics[width=0.8\linewidth]{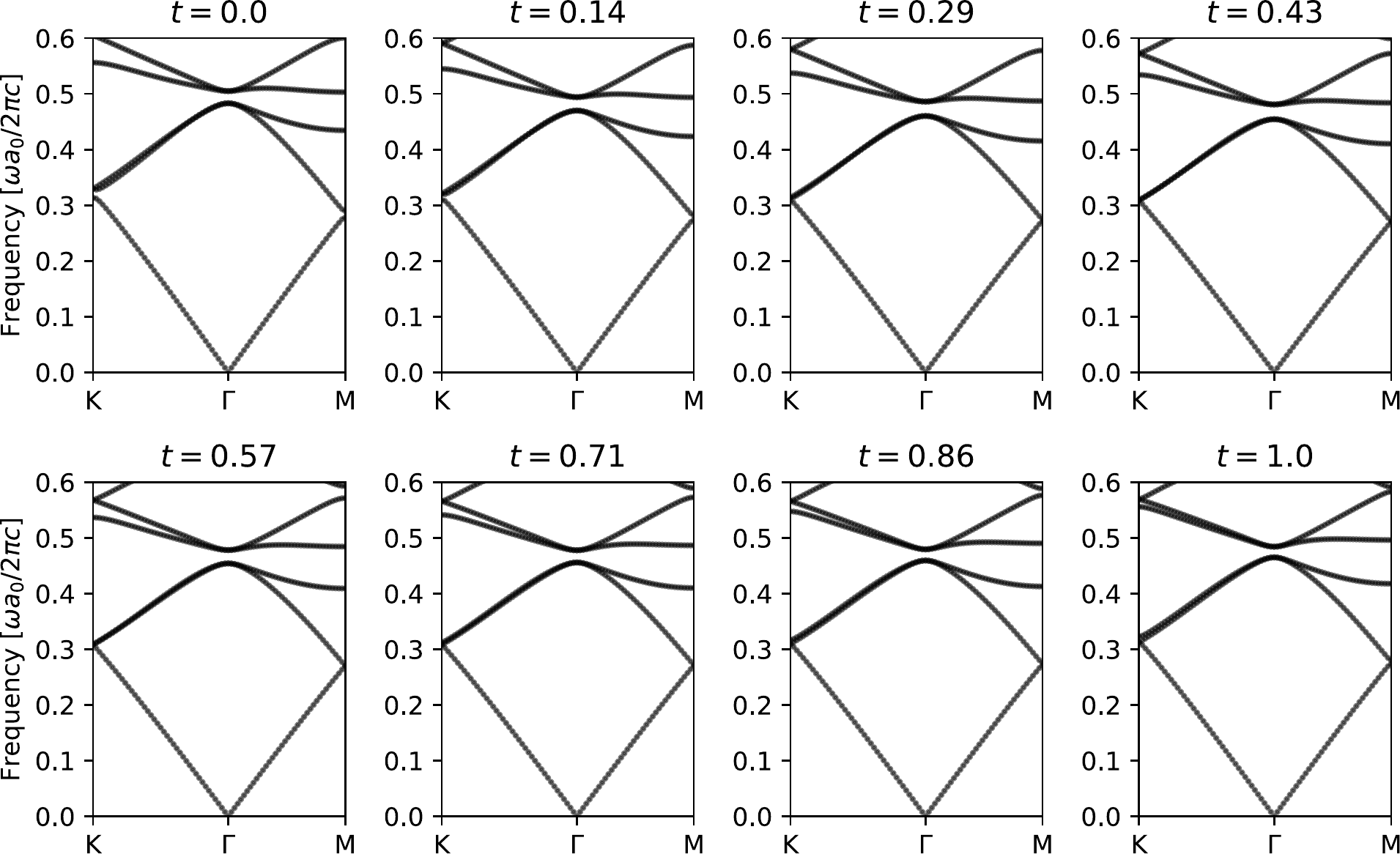}
	\caption{Continuous deformation between ($t=0$) the fragilely topological crystal of Blanco de Paz et al \cite{blanco2019engineering} and ($t=1$) the topological crystal of Wu and Hu \cite{wu2015scheme}.}
	\label{fig:fragile-OAL-deformation}
\end{figure}

Figure \ref{fig:fragile-OAL-deformation} shows the band structures of crystals as we deform between a fragilely topological crystal \cite{blanco2019engineering} and the obstructed atomic limit of Wu and Hu \cite{wu2015scheme},
\begin{align}
	R & =(1-t)R^{\mathrm{frag}}+tR^{\mathrm{OAL}}\\
	d_{1} & =(1-t)d_{1}^{\mathrm{frag}}+td_{1}^{\mathrm{OAL}}\\
	d_{2} & =(1-t)d_{2}^{\mathrm{frag}}+td_{2}^{\mathrm{OAL}}
\end{align}
where $a_{0}/R^{\mathrm{frag}}=2.9$, $d_{1}^{\mathrm{frag}}=0.4a_{0}$, $d_{2}^{\mathrm{frag}}=0.13a_{0}$, and $a_{0}/R^{\mathrm{OAL}}=2.9$, $d_{1}^{\mathrm{OAL}}=d_{2}^{\mathrm{OAL}}=2R^{\mathrm{OAL}}/3$. The deformation preserves all symmetries of the crystal, and the valence bands (first three bands) remain isolated from the conduction bands throughout the deformation.

%\clearpage

\end{document}